\documentclass[aps,pra,preprint,amsmath]{revtex4-1}
\usepackage{silence}
\usepackage{mathrsfs}
\usepackage{graphicx}
\usepackage{epstopdf}
\usepackage{amsthm}
\usepackage{amsmath}
\usepackage{amsfonts}
\usepackage{amssymb}
\usepackage{braket}
\usepackage{bbold}
\usepackage{subfigure}
\usepackage{appendix}
\usepackage{tikz}
\usepackage{bm}
\usepackage{float}
\def\be{\begin{equation}}
\def\ee{\end{equation}}
\def\ba{\begin{array}}
\def\ea{\end{array}}

\def\1{{\bf{1}}}

\input amssym.def
\begin{document}
\title{\bf Uncertainty of quantum channels based on symmetrized $\rho$-absolute variance and modified Wigner-Yanase skew information}

\author{Cong Xu$^1$}
\author{Qing-Hua Zhang$^2$}
\author{Shao-Ming Fei$^{1,3}$}
\email{Cong Xu\\2230501028@cnu.edn.cn\\Shao-Ming Fei\\feishm@cnu.edu.cn}
\affiliation{
$^1$School of Mathematical Sciences, Capital Normal University, Beijing 100048, China\\
$^2$School of Mathematics and Statistics, Changsha University of Science and Technology, Changsha 410114, China\\
$^3$Max-Planck-Institute for Mathematics in the Sciences, 04103 Leipzig, Germany
}

\begin{abstract}
We present the uncertainty relations in terms of the symmetrized $\rho$-absolute variance, which generalizes the uncertainty  relations for arbitrary operator (not necessarily Hermitian) to quantum channels. By recalling the quantity $|U_{\rho}|(\Phi)$ proposed by Zhang $et \  al.$ (Quantum Inf. Process. \textbf{22} 456, 2023), which involves terms of more quantum mechanical nature. We also establish the tighter uncertainty relations for quantum channels by using Cauchy-Schwarz inequality. Detailed examples are provided to illustrate the tightness of our results.

\end{abstract}

\maketitle

\noindent {\bf 1. Introduction}

As one of the cornerstones of quantum mechanics, uncertainty principle has been widespreadly concerned since proposed by Heisenberg \cite{HW} in $1927$.
The standard form of Heisenberg uncertainty relation was given by two arbitrary observables $A$ and $B$ with respect to a quantum state $\rho$ \cite{RH},
\begin{align}\label{eq1}
 V_{\rho}(A)V_{\rho}(B)\geq \frac{1}{4}|\mathrm{Tr}(\rho[A,B])|^2,
\end{align}
where $V_{\rho}(A)=\mathrm{Tr}(\rho A^2)-(\mathrm{Tr}\rho A)^2$ is the conventional variance and $[A,B]=AB-BA$ denotes the commutator. Later (\ref{eq1}) was refined by Schr\"{o}dinger \cite{SE,SE2},
\begin{align}\label{eq2}
 V_{\rho}(A)V_{\rho}(B)\geq \frac{1}{4}|\mathrm{Tr}(\rho[A,B])|^2+\frac{1}{4}|\mathrm{Tr}(\rho\{A_0,B_0\})|^2,
\end{align}
where $A_0=A-\mathrm{Tr}(\rho A)$, $B_0=B-\mathrm{Tr}(\rho B)$ and $\{A_0,B_0\}=A_0B_0+A_0B_0$ denotes the anti-commutator.

The quantities $V_{\rho}(A)$, $V_{\rho}(B)$ and $\mathrm{Tr}(\rho[A,B])$ are all purely quantum terms when $\rho$ is a pure state. If $\rho$ is a mixed state, inequality (\ref{eq1}) is intrinsically of a mixed nature since the quantity $V_{\rho}(A)$ is a hybrid of both classical mixing and quantum uncertainty \cite{LUO1}. Luo \cite{LUO8} striped certain classical mixing uncertainty off the variance and introduced the quantity $U_{\rho}(A)=\sqrt{V^2_{\rho}(A)-[V_{\rho}(A)-I_{\rho}(A)]^2}$, where $I_{\rho}(A)=-\frac{1}{2}\mathrm{Tr}
\left(\left[\sqrt{\rho},A\right]^2\right)$ is the Wigner-Yanase (WY) skew information \cite{WY}. $U_{\rho}(A)$ satisfies the following uncertainty relation \cite{LUO8},
\begin{align}\label{eq3}
 U_{\rho}(A)U_{\rho}(B)\geq \frac{1}{4}|\mathrm{Tr}(\rho[A,B])|^2.
\end{align}

As the most general description of quantum measurement, quantum channels play an essential role in quantum theory \cite{NC}. For a quantum channel $\Phi$ with Kraus representation $\Phi(\rho)=\sum_{i}E_i\rho E_i^\dag$, where $\sum_{i}E_i^\dag E_i=I$, with $I$ the identity operator and a quantum state $\rho$. Luo $et \ al.$ \cite{LUO9} defined the quantity $|I_{\rho}|(\Phi)=\sum_i|I_{\rho}|(E_i)$, the sum of the uncertainties of the Kraus operators $\{E_i\}$, to characterize the uncertainty relations of a quantum state $\rho$ with respect to the quantum channel $\Phi$, which can be also used to quantity the coherence in a more broad context.  Based on the modified Wigner-Yanase (MWY) skew information, $|I_{\rho}|(K)=\frac{1}{2}\mathrm{Tr}\left([\sqrt{\rho}
,K]^{\dag}[\sqrt{\rho},K]\right)$\cite{DD1}, where the operator $K$ not necessarily Hermitian,
Fu $et \ al.$ \cite{FSS} introduced the summation form uncertainty relations of two quantum channels. The product form uncertainty relations have been also explored \cite{ZN}. By using a sequence of "fine-grained" inequalities, Hu $et \ al.$ \cite{HJ1} improved the lower bound of the Theorem 1 in Ref. \cite{ZN}. Recently, the product and summation forms of uncertainty relations for quantum channels and observables have also been studied intensely \cite{XWF2,XWF1,SL,XWF3,XWF4,ZL,ZWF1,YANA1,YANA2,FSS,ZWF2,CAL,RRNL,LUO,HLTG,ZN,HJ1,ZL,ZWF4,ZWF5,WZF,HJ2,MZF,HJ}.


The remainder of this paper is structured as follows. In Section 2, we recall some
basic concepts and related results. In Section 3, we investigate the uncertainty relations for quantum channels based on the symmetrized $\rho$-absolute variance. The tighter product and summation forms of the uncertainty relations for quantum channels in terms of MWY skew information are also explored. Besides, we compare our results with the existing ones by detailed examples. We conclude and summarize in Section 4.

\noindent {\bf 2. Preliminaries}

Let $\mathcal{H}$ be an $n$-dimensional Hilbert space. Denote $\mathcal{B(H)}$ and $\mathcal{S(H)}$ the sets of
all bounded linear operators and Hermitian operators on $\mathcal{H}$, respectively. For $K$, $L\in\mathcal{B(H)}$, $\langle K,L\rangle=\mathrm{Tr}(K^\dag L)$ is the inner product. The norm of $K\in\mathcal{B(H)}$ is defined by $\|K\|=\sqrt{\mathrm{Tr}(K^{\dag}K)}$.

{\bf Definition 1} For a quantum state $\rho\in\mathcal{D(H)}$ and an operator $K\in\mathcal{B(H)}$, the $\rho$-absolute variance of $K$ is defined by \cite{Gudder}
\begin{align}\label{eq4}
 |V_{\rho}|(K):=\mathrm{Tr}({\rho}|K_0|^2)=\|K_0\sqrt{\rho}\|^{2}=\mathrm{Tr}({\rho}|K|^2)-|\mathrm{Tr}({\rho}K)|^2,
\end{align}
where $K_0=K-\mathrm{Tr}(\rho K)$ and $|K|=(K^{\dag}K)^\frac{1}{2}\in\mathcal{S(H)}$. The symmetrized $\rho$-absolute variance is defined by \cite{DD1}
\begin{align}\label{eq5}
 |V^\circ_{\rho}|(K)=\frac{1}{2}[|V_{\rho}|(K)+|V_{\rho}|(K^{\dag})],
\end{align}
where $|V_{\rho}|(K)$ is the $\rho$-absolute variance. Note that both $|V_{\rho}|(K)$ and $|V^\circ_{\rho}|(K)$ reduce to the conventional variance $V_{\rho}(K)=\mathrm{Tr}(\rho K^2)-(\mathrm{Tr}\rho K)^2$ when $K$ is a Hermitian operator.

{\bf Definition 2} The symmetrized commutator $[X,Y]^\circ$ of $X$ and $Y\in\mathcal{B(H)}$ is defined by \cite{DD1}
\begin{align}\label{eq6}
 [X,Y]^\circ=\frac{1}{2}([X,Y]+[X^{\dag},Y^{\dag}]),
\end{align}
and the symmetrized anti-commutator $\{X,Y\}^\circ$ of $X$ and $Y$ is defined by \cite{DD1}
\begin{align}\label{eq7}
 \{X,Y\}^\circ=\frac{1}{2}(\{X,Y\}+\{X^{\dag},Y^{\dag}\}).
\end{align}

{\bf Lemma 1} For any $K\in\mathcal{B(H)}$ and state $\rho$, if $K=A+\mathrm{i}B$ is the Cartesian decomposition of $K$, i.e., $A,B\in\mathcal{S(H)}$, then \cite{DD1}
\begin{align}\label{eq8}
|\mathrm{Tr}(\rho K)|^2=|\mathrm{Tr}(\rho A)|^2+|\mathrm{Tr}(\rho B)|^2.
\end{align}

Based on Definition 1 and Definition 2,
Dou $et \ al.$ \cite{DD1} generalized the inequalitie (\ref{eq1}-\ref{eq3}) to the case arbitrary operators $K$ and $L\in\mathcal{B(H)}$,
\begin{align}\label{eq9}
 |V^\circ_{\rho}|(K)|V^\circ_{\rho}(L)|\geq \frac{1}{4}|\mathrm{Tr}(\rho[K,L])|^2,
\end{align}
\begin{align}\label{eq10}
 |V^\circ_{\rho}|(K)|V^\circ_{\rho}(L)|\geq \frac{1}{4}|\mathrm{Tr}(\rho[K,L]^\circ)|^2+\frac{1}{4}|\mathrm{Tr}(\rho\{K,L\}^\circ)|^2
\end{align}
and
\begin{align}\label{eq11}
 |U_{\rho}|(K)|U_{\rho}(L)|\geq \frac{1}{4}|\mathrm{Tr}(\rho[K,L]^\circ)|^2,
\end{align}
where $|U_{\rho}|(K)=\sqrt{|V^\circ_{\rho}|^2(K)-(|V^\circ_{\rho}|(K)-|I_{\rho}|(K))^2}$, $[K,L]^\circ$ and $\{K,L\}^\circ$ are the symmetrized commutator and the symmetrized anti-commutator, respectively \cite{DD1}. The inequalities (\ref{eq9})-(\ref{eq11}) give rise to the inequalities (\ref{eq1})-(\ref{eq3}) when $K,L\in\mathcal{S(H)}$.

Following the idea in Ref. \cite{LUO9}, Sun $et \ al.$  defined the quantity $|V^\circ_{\rho}|(\Phi)=\sum_i|V^\circ_{\rho}|(E_i)$, the sum of the uncertainties of the
Kraus operators $\{E_i\}$ \cite{SL}. The quantities $|V^\circ_{\rho}|(\Phi)$ and $|I_{\rho}|(\Phi)$ characterize the total uncertainty and the quantum uncertainty of the quantum channel $\Phi$, respectively \cite{ZWF4,SL}. Zhang $et \ al.$ \cite{ZWF4} further extended the quantity $|U_{\rho}|(K)$ to
\begin{align}\label{eq12}
 |U_{\rho}|(\Phi)=&\sqrt{|V^\circ_{\rho}|^2(\Phi)-[|V^\circ_{\rho}|(\Phi)-
|I_{\rho}|(\Phi)]^2}\notag\\
=&\sqrt{|V^\circ_{\rho}|^2(\Phi)-|C_{\rho}|^2(\Phi)} \notag\\
=&\sqrt{[|V^\circ_{\rho}|(\Phi)+|C_{\rho}|(\Phi)][|V^\circ_{\rho}|(\Phi)-\Phi|C_{\rho}|(\Phi)]} \notag\\
=&\sqrt{|\widetilde{I}|_{\rho}(\Phi)|\widetilde{J}|_{\rho}(\Phi)}.
\end{align}
By direct verification we have
$|\widetilde{I}|_{\rho}(\Phi)=\sum_i|I_{\rho}|(E_{i0})=\frac{1}{2}\sum_i\mathrm{Tr}\left([\sqrt{\rho}
,E_{i0}]^{\dag}[\sqrt{\rho},E_{i0}]\right)$ and
$|\widetilde{J}|_{\rho}(\Phi)=\sum_i|J_{\rho}|(E_{i0})=\frac{1}{2}\sum_i\mathrm{Tr}\left(\{\sqrt{\rho}
,E_{i0}\}^{\dag}\{\sqrt{\rho},E_{i0}\}\right)$ where $E_{i0}=E_i-\mathrm{Tr}(\rho E_i)$.

For two arbitrary quantum channels $\Phi$ and $\Psi$, it has been proved that \cite{ZWF4}
\begin{align}\label{eq13}
|U_{\rho}|(\Phi)|U_{\rho}|(\Psi)\geq\frac{1}{4}\sum_{ij}|\mathrm{Tr}([F_j,{E^{\dag}_i}]\rho)|^2
\end{align}
and
\begin{align}\label{eq14}
|U_{\rho}|^2(\Psi)+|U_{\rho}|^2(\Phi)\geq\frac{1}{2}\sum_{ij}\left|\langle[\sqrt{\rho},F_i],[\sqrt{\rho},E_i]\rangle \left(\langle\{\sqrt{\rho},F_j\},\{\sqrt{\rho},E_j\}\rangle-4\langle F_j^{\dag}\rangle\langle E_j\rangle \right)\right|,
\end{align}
where $\Phi(\rho)=\sum_{i}E_i\rho E_i^\dag$ and $\Psi(\rho)=\sum_{j}F_j\rho F_j^\dag$.

\noindent {\bf 3. Uncertainty relations of quantum channels in terms of the symmetrized $\rho$-absolute variance and MWY skew information}

In this section, we generalize the uncertainty relations (\ref{eq9})-(\ref{eq11}) of arbitrary operators to quantum channels.

{\bf Theorem 1} For two arbitrary quantum channels with Kraus representations $\Phi(\rho)=\sum_{i=1}^{N}E_i\rho E_i^\dag$ and $\Psi(\rho)=\sum_{j=1}^{N}F_j\rho F_j^\dag$ on an $n$-dimensional Hilbert space $\mathcal{H}$, the following uncertainty relations hold,
\begin{align}\label{eq15}
|V^\circ_{\rho}|(\Phi)|V^\circ_{\rho}|(\Psi)\geq\mathrm{max}\left\{\frac{1}{4N^2}\left|\sum_{ij}\mathrm{Tr}(\rho[E_i,F_j])\right|^2,\frac{1}{4N^2}\left|\sum_{ij}\mathrm{Tr}(\rho\{E_{i0},F_{j0}\})\right|^2\right\}. \end{align}

{\it Proof}
Define $2\times2$ matrices $\mathcal{A}_{ij}$ and $\mathcal{B}_{ij}$ by
\begin{equation*}
\mathcal{A}_{ij}=
\begin{pmatrix}
|V_{\rho}|(E^{\dag}_{i0})& \langle\sqrt{\rho}, E_{i0}F_{j0}\sqrt{\rho}\rangle& \\
\langle\sqrt{\rho}, F^{\dag}_{j0}E^{\dag}_{i0}\sqrt{\rho}\rangle& |V_{\rho}|(F_{j0})& \\
\end{pmatrix}
\end{equation*}
\\
and
\begin{equation*}
\mathcal{B}_{ij}=
\begin{pmatrix}
|V_{\rho}|(E_{i0})& \langle\sqrt{\rho}, -F_{j0}E_{i0}\sqrt{\rho}\rangle& \\
\langle\sqrt{\rho}, -E^{\dag}_{i0}F^{\dag}_{j0}\sqrt{\rho}\rangle& |V_{\rho}|(F^{\dag}_{j0})& \\
\end{pmatrix},
\end{equation*}
respectively. It is easily seen that $\overline{\langle\sqrt{\rho}, F^{\dag}_{j0}E^{\dag}_{i0}\sqrt{\rho}\rangle}=\langle\sqrt{\rho}, E_{i0}F_{j0}\sqrt{\rho}\rangle=\langle E^{\dag}_{i0}\sqrt{\rho}, F_{j0}\sqrt{\rho}\rangle$ and $|V_{\rho}|(E^{\dag}_{i0})|V_{\rho}|(F_{j0})
\geq|\langle E^{\dag}_{i0}\sqrt{\rho}, F_{j0}\sqrt{\rho}\rangle|^2=
|\langle\sqrt{\rho}, E_{i0}F_{j0}\sqrt{\rho}\rangle|^2$ by Cauchy-Schwarz inequality.
Hence, $\mathcal{A}_{ij}$ and $\sum_{ij}\mathcal{A}_{ij}$ are positive semi-definite. In a similar way, we can verify that the matrices $\mathcal{B}_{ij}$, $\sum_{ij}\mathcal{B}_{ij}$ and $\sum_{ij}(\mathcal{A}_{ij}+\mathcal{B}_{ij})$ are all positive semi-definite. Moreover, by using the fact that
$|V^\circ_{\rho}|(E_{i0})=|V^\circ_{\rho}|(E_i)$,
$[E_{i0},F_{j0}]=[E_i,F_j]$ and (\ref{eq5}), $\sum_{ij}(\mathcal{A}_{ij}+\mathcal{B}_{ij})$ can be rewritten as
\begin{equation*}
\sum_{ij}(\mathcal{A}_{ij}+\mathcal{B}_{ij})=
\begin{pmatrix}
N\sum_{i}[|V_{\rho}|(E^{\dag}_{i0})+|V_{\rho}|(E_{i0})]& \sum_{ij}\langle\sqrt{\rho}, (E_{i0}F_{j0}-F_{j0}E_{i0})\sqrt{\rho}\rangle& \\
\sum_{ij}\langle\sqrt{\rho}, (F^{\dag}_{j0}E^{\dag}_{i0}-E^{\dag}_{i0}F^{\dag}_{j0})\sqrt{\rho}\rangle& N\sum_{j}[|V_{\rho}|(F_{j0})+|V_{\rho}|(F^{\dag}_{j0})]& \\
\end{pmatrix}
\end{equation*}
\begin{equation*}
=\begin{pmatrix}
2N|V^\circ_{\rho}|(\Phi)& \sum_{ij}\langle\sqrt{\rho}, [E_{i},F_{j}]\sqrt{\rho}\rangle& \\
\sum_{ij}\langle\sqrt{\rho}, [E_i,F_j]^{\dag}\sqrt{\rho}\rangle& 2N|V^\circ_{\rho}|(\Psi)& \\
\end{pmatrix}.
\end{equation*}
Therefore, we have
\begin{align*}
|V^\circ_{\rho}|(\Phi)|V^\circ_{\rho}|(\Psi)\geq\frac{1}{4N^2}\left|\sum_{ij}\mathrm{Tr}(\rho[E_i,F_j])\right|^2. \end{align*}

In a similar way, we prove that the following $2\times2$ matrix $\mathcal{C}_{ij}$,
\begin{equation*}
\mathcal{C}_{ij}=
\begin{pmatrix}
|V_{\rho}|(E_{i0})& \langle\sqrt{\rho}, F_{j0}E_{i0}\sqrt{\rho}\rangle& \\
\langle\sqrt{\rho}, E^{\dag}_{i0}F^{\dag}_{j0}\sqrt{\rho}\rangle& |V_{\rho}|(F^{\dag}_{j0})& \\
\end{pmatrix}
\end{equation*}
is positive semi-definite too. It is easily seen $\sum_{ij}(\mathcal{A}_{ij}+\mathcal{C}_{ij})\geq0$. Therefore,
\begin{align*}
|V^\circ_{\rho}|(\Phi)|V^\circ_{\rho}|(\Psi)\geq\frac{1}{4N^2}
\left|\sum_{ij}\mathrm{Tr}(\rho\{E_{i0},F_{j0}\})\right|^2.
\end{align*}
$\Box$

{\bf Theorem 2} For two arbitrary quantum channels given by $\Phi(\rho)=\sum_{i=1}^{N}E_i\rho E_i^\dag$ and $\Psi(\rho)=\sum_{j=1}^{N}F_j\rho F_j^\dag$, we have
\begin{align}\label{eq16}
|V^\circ_{\rho}|(\Phi)|V^\circ_{\rho}|(\Psi)\geq\frac{1}{4N^2}
\left(\left|\sum_{ij}\mathrm{Tr}(\rho\{E_{i0},F_{j0}\}^\circ)\right|^2
+\left|\sum_{ij}\mathrm{Tr}(\rho[E_{i0},F_{j0}]^\circ)\right|^2\right).
\end{align}

{\it Proof}
Define $2\times2$ matrices $\mathcal{A}_{ij}$ and $\mathcal{D}_{ij}$ by
\begin{equation*}
\mathcal{A}_{ij}=
\begin{pmatrix}
|V_{\rho}|(E^{\dag}_{i0})& \langle\sqrt{\rho}, E_{i0}F_{j0}\sqrt{\rho}\rangle& \\
\langle\sqrt{\rho}, F^{\dag}_{j0}E^{\dag}_{i0}\sqrt{\rho}\rangle& |V_{\rho}|(F_{j0})& \\
\end{pmatrix}
\end{equation*}
\\
and
\begin{equation*}
\mathcal{D}_{ij}=
\begin{pmatrix}
|V_{\rho}|(E_{i0})& \langle\sqrt{\rho}, E^{\dag}_{i0}F^{\dag}_{j0}\sqrt{\rho}\rangle& \\
\langle\sqrt{\rho}, F_{j0}E_{i0}\sqrt{\rho}\rangle& |V_{\rho}|(F^{\dag}_{j0})& \\
\end{pmatrix},
\end{equation*}
respectively. Similar to the proof of Theorem 1, it can be verified that $\sum_{ij}\mathcal{A}_{ij}$,  $\sum_{ij}\mathcal{D}_{ij}$ and $\sum_{ij}(\mathcal{A}_{ij}+\mathcal{D}_{ij})$ are positive semi-definite. Moreover,
\begin{equation*}
\sum_{ij}(\mathcal{A}_{ij}+\mathcal{D}_{ij})=
\begin{pmatrix}
N\sum_{i}(|V_{\rho}|(E^{\dag}_{i0})+|V_{\rho}|(E_{i0}))& \sum_{ij}\langle\sqrt{\rho}, (E_{i0}F_{j0}\sqrt{\rho}+E^{\dag}_{i0}F^{\dag}_{j0})\sqrt{\rho}\rangle& \\
\sum_{ij}\langle\sqrt{\rho}, (F^{\dag}_{j0}E^{\dag}_{i0}\sqrt{\rho}+F^{\dag}_{j0}E^{\dag}_{i0})\sqrt{\rho}\rangle& N\sum_{j}(|V_{\rho}|(F_{j0})+|V_{\rho}|(F^{\dag}_{j0}))& \\
\end{pmatrix}
\end{equation*}
\begin{equation*}
=\begin{pmatrix}
2N|V^\circ_{\rho}|(\Phi)& \sum_{ij}\langle\sqrt{\rho}, (E_{i0}F_{j0}+E^{\dag}_{i0}F^{\dag}_{j0})\sqrt{\rho}\rangle& \\
\sum_{ij}\langle\sqrt{\rho}, (F^{\dag}_{j0}E^{\dag}_{i0}+F_{j0}E_{i0})\sqrt{\rho}\rangle& 2N|V^\circ_{\rho}|(\Psi)& \\
\end{pmatrix}.
\end{equation*}
In this case, $E_{i0}F_{j0}+E^{\dag}_{i0}F^{\dag}_{j0}$ can be rewritten as
\begin{align*}
E_{i0}F_{j0}+E^{\dag}_{i0}F^{\dag}_{j0}\notag
=&\frac{1}{2}([E_{i0},F_{j0}]+[E^{\dag}_{i0},F^{\dag}_{j0}])
+\frac{1}{2}(\{E_{i0},F_{j0}\}+\{E^{\dag}_{i0},F^{\dag}_{j0}\}) \notag\\
=&\mathrm{i}\frac{1}{2}(-\mathrm{i}([E_{i0},F_{j0}]+[E^{\dag}_{i0},F^{\dag}_{j0}]))+\frac{1}{2}(\{E_{i0},F_{j0}\}+\{E^{\dag}_{i0},F^{\dag}_{j0}\})\notag\\
=&\mathrm{i}(-\mathrm{i}[E_{i0},F_{j0}]^\circ)+\{E_{i0},F_{j0}\}^\circ.
\end{align*}
Hence,
\begin{align*}
|V^\circ_{\rho}|(\Phi)|V^\circ_{\rho}|(\Psi)\notag
\geq\frac{1}{4N^2}&\left|\sum_{ij}\langle\sqrt{\rho}, [\mathrm{i}(-\mathrm{i}[E_{i0},F_{j0}]^\circ)+\{E_{i0},F_{j0}\}^\circ]\sqrt{\rho}\rangle\right|^2 \notag\\
=\frac{1}{4N^2}&\left|\sum_{ij}\mathrm{Tr}(\rho\{E_{i0},F_{j0}\}^\circ)+\mathrm{i}\sum_{ij}\mathrm{Tr}\rho(-\mathrm{i}[E_{i0},F_{j0}]^\circ)\right|^2
\notag\\
=\frac{1}{4N^2}&\left(\left|\sum_{ij}\mathrm{Tr}(\rho\{E_{i0},F_{j0}\}^\circ)\right|^2+\left|\sum_{ij}\mathrm{Tr}\rho([E_{i0},F_{j0}]^\circ)\right|^2\right),
\end{align*}
where the last equality follows from Lemma 1.

The lower bounds given in Theorem 1 and 2 are generally different for different channels $\Phi$ and $\Psi$ and states $\rho$. They are complementary in lower bounding the product $|V^\circ_{\rho}|(\Phi)|V^\circ_{\rho}|(\Psi)$.

Next we present tighter uncertainty relations of (\ref{eq13}) and (\ref{eq14}) for quantum channels based on MWY skew information. Let $\Phi$ and $\Psi$ be two arbitrary quantum channels on $n$-dimensional Hilbert space $\mathcal{H}$ with Kraus representations $\Phi(\rho)=\sum_{i}E_i\rho E_i^\dag$ and $\Psi(\rho)=\sum_{j}F_j\rho F_j^\dag$. Denote
$|\rho^{E_{i0}}_k\rangle=\rho^{E_{i0}}|k\rangle=[\sqrt{\rho},E_{i0}]|k\rangle$, $|\rho^{F_{j0}}_l\rangle=\rho^{F_{j0}}|l\rangle=\{\sqrt{\rho},F_{j0}\}|l\rangle$, $|\rho^{\widetilde{E}_{i0}}_k\rangle
=\rho^{\widetilde{E}_{i0}}|k\rangle=\{\sqrt{\rho},E_{i0}\}|k\rangle$ and $|\rho^{\widetilde{F}_{j0}}_l\rangle
=\rho^{\widetilde{F}_{j0}}|l\rangle=[\sqrt{\rho},F_{j0}]|l\rangle$, where $\{|t\rangle\}^n_{t=1}$ is an orthonormal basis in $\mathcal{H}$. Denote
$\vec{e}_n^i=(|\rho_1^{E_{i0}}\rangle,|\rho_2^{E_{i0}}\rangle\cdots,|\rho_n^{E_{i0}}\rangle)^T$,
$\vec{f}_n^j=(|\rho_1^{{F}_{j0}}\rangle,|\rho_2^{{F}_{j0}}\rangle\cdots,|\rho_n^{{F}_{j0}}\rangle)^T$,
$\vec{g}_n^i=(|\rho_1^{\widetilde{E}_{i0}}\rangle,
|\rho_2^{\widetilde{E}_{i0}}\rangle\cdots,|\rho_n^{\widetilde{E}_{i0}}\rangle)^T$,
$\vec{h}_n^j=(|\rho_1^{\widetilde{F}_{j0}}\rangle,|\rho_2^{\widetilde{F}_{j0}}\rangle\cdots,
|\rho_n^{\widetilde{F}_{j0}}\rangle)^T$,
and $\vec{e}_1^i=(|\rho_1^{E_{i0}}\rangle,\cdots,\cdots,\vec{0}_{n\times 1})^T$,
$\vec{f}_1^j=(|\rho_1^{F_{j0}}\rangle,\cdots,\vec{0}_{n\times 1})^T$,
$\vec{g}_1^i=(|\rho_1^{\widetilde{E}_{i0}}\rangle,\cdots,\vec{0}_{n\times 1})^T$,
$\vec{h}_1^j=(|\rho_1^{\widetilde{F}_{j0}}\rangle,\cdots,\vec{0}_{n\times 1})^T$.
In particular, we have $\vec{e}_0^i=\vec{f}_0^j=\vec{g}_0^i=\vec{h}_0^j=\vec{0}^T_{n^2\times 1}$. $\vec{e}_{1,c}$, $\vec{f}_{1,c}$, $\vec{g}_{1,c}$ and $\vec{h}_{1,c}$ are the complementary vectors of $\vec{e}_1^i$, $\vec{f}_1^j$, $\vec{g}_1^i$ and $\vec{h}_1^j$, respectively. With these notions $|I_{\rho}|(E_{i0})$ and $|J_{\rho}|(F_{j0})$ can be written as
\begin{align*}
|I_{\rho}|(E_{i0})=&\frac{1}{2}\mathrm{Tr}\left([\sqrt{\rho}
,E_{i0}]^{\dag}[\sqrt{\rho},E_{i0}]\right)\notag
=\frac{1}{2}\left(\sum_{k=1}^n\langle k|(\rho^{E_{i0}})^{\dag}\rho^{E_{i0}}|k\rangle\right)\notag\\
=&\frac{1}{2}\left(\sum_{k=1}^n \||\rho^{E_{i0}}_k\rangle\|^2\right)
=\frac{1}{2}\|\vec{e}_n^i\|^2
\end{align*}
and
\begin{align*}
|J_{\rho}|(F_{j0})=&\frac{1}{2}\mathrm{Tr}\left(\{\sqrt{\rho}
,F_{j0}\}^{\dag}\{\sqrt{\rho},F_{j0}\}\right)\notag
=\frac{1}{2}\left(\sum_{l=1}^n\langle
l|(\rho^{F_{j0}})^{\dag}\rho^{F_{j0}}|l\rangle\right)\notag\\
=&\frac{1}{2}\left(\sum_{l=1}^n \||\rho^{F_{j0}}_l\rangle\|^2\right)
=\frac{1}{2}\|\vec{f}_n^j\|^2.
\end{align*}
The quantities $|I_{\rho}|(F_{j0})=\frac{1}{2}\|\vec{h}_n^j\|^2$ and $|J_{\rho}|(E_{i0})=\frac{1}{2}\|\vec{g}_n^i\|^2$ are defined similarly. We have the following theorems.

{\bf Theorem 3} For two arbitrary quantum channels with Kraus representations $\Phi(\rho)=\sum_{i}^NE_i\rho E_i^\dag$ and $\Psi(\rho)=\sum_{j}^NF_j\rho F_j^\dag$ on an $n$-dimensional Hilbert space $\mathcal{H}$, the following uncertainty relation holds,
\begin{align}\label{eq21}
|U_{\rho}|(\Phi)|U_{\rho}|(\Psi)\geq\sqrt{I_1\widetilde{I}_1},
\end{align}
where $I_1=\sum_{ij}I_1^{i,j}$ and $\widetilde{I}_1=\sum_{ij}\widetilde{I}_1^{i,j}$.

{\it Proof}  By using the Cauchy-Schwarz inequality, we have
\begin{align}\label{eq22}
|I_{\rho}|({E_{i0}})|J_{\rho}|({F_{j0}})=&\frac{1}{4}\|\vec{e}_n^i\|^2\|\vec{f}_n^j\|^2\notag\\
=&\frac{1}{4}(\|\vec{e}_1^i\|^2+\|\vec{e}_{1,c}^i\|^2)(\|\vec{f}_1^j\|^2+\|\vec{f}_{1,c}^j\|^2) \notag\\
=&\frac{1}{4}\left(\|\vec{e}_1^i\|^2\|\vec{f}_1^j\|^2+\|\vec{e}_1^i\|^2\|\vec{f}_{1,c}^j\|^2+\|\vec{e}_{1,c}^i\|^2(\|\vec{f}_1^j\|^2+\|\vec{f}_{1,c}^j\|^2)\right)\notag\\
\geq& \frac{1}{4}\left(|\langle\vec{e}_1^i,\vec{f}_1^j\rangle|^2+\|\vec{e}_1^i\|^2\|\vec{f}_{1,c}^j\|^2+\|\vec{e}_{1,c}^i\|^2(\|\vec{f}_1^j\|^2+\|\vec{f}_{1,c}^j\|^2)\right).
\end{align}
Denote $I_0^{i,j}=|I_{\rho}|({E_{i0}})|J_{\rho}|({F_{j0}})$ and
$I_1^{i,j}=\frac{1}{4}\left(|\langle\vec{e}_1^i,\vec{f}_1^j\rangle|^2+\|\vec{e}_1^i\|^2\|\vec{f}_{1,c}^j\|^2+\|\vec{e}_{1,c}^i\|^2(\|\vec{f}_1^j\|^2
\right.
\nonumber\\
\left.+\|\vec{f}_{1,c}^j\|^2)\right)$.
We obtain
\begin{align}\label{eq23}
I_1^{i,j}=&I_0^{i,j}-\frac{1}{4}(\|\vec{e}_1^i\|^2\|\vec{f}_1^j\|^2-|\langle\vec{e}_1^i,\vec{f}_1^j\rangle|^2) \notag\\
=&|I_{\rho}|({E_{i0}})|J_{\rho}|({F_{j0}})-\frac{1}{4}\left(\langle\langle\rho_1^{E_{i0}}|,|\rho_1^{E_{i0}}\rangle\rangle\langle\langle\rho_1^{F_{j0}}|,|\rho_1^{F_{j0}}\rangle\rangle-|\langle\langle\rho_1^{E_{i0}}|,|\rho_1^{F_{j0}}\rangle\rangle|^2\right)\notag\\
=&|I_{\rho}|({E_{i0}})|J_{\rho}|({F_{j0}})-\frac{1}{4}(\langle 1|[\sqrt{\rho},E_{i0}]^{\dag}[\sqrt{\rho},E_{i0}]|1\rangle\langle1|\{\sqrt{\rho},F_{j0}\}^{\dag}\{\sqrt{\rho},F_{j0}\}|1\rangle) \notag\\
+&\frac{1}{4}(|\langle 1|[\sqrt{\rho},E_{i0}]^{\dag}\{\sqrt{\rho},F_{j0}\}|1\rangle|^2).
\end{align}

Summing over the indices $i$ and $j$ on both sides of the inequality (\ref{eq22}), we have
$\widetilde{I}_{\rho}(\Phi)\widetilde{J}_{\rho}(\Psi)\geq I_1$, which
is saturated if and only if $|\rho_1^{E_{i0}}\rangle$ and $|\rho_1^{F_{j0}}\rangle$ are linear dependent for arbitrary $i$ and $j$.

Similarly, for $|I_{\rho}|({F_{j0}})|J_{\rho}|({E_{i0}})$, we have $\widetilde{I}_1^{i,j}=\frac{1}{4}(|\langle\vec{h}_1^j,\vec{g}_1^i\rangle|^2
+\|\vec{h}_1^j\|^2\|\vec{g}_{1,c}^i\|^2+\|\vec{h}_{1,c}^j\|^2(\|\vec{g}_1^i\|^2
+\|\vec{g}_{1,c}^i\|^2))$ and
$\widetilde{I}_0^{i,j}=|I_{\rho}|({F_{j0}})|J_{\rho}|({E_{i0}})$. We obtain
 \begin{align}\label{eq25}
\widetilde{I}_1^{i,j}=&\widetilde{I}_0^{i,j}-\frac{1}{4}(\|\vec{h}_1^j\|^2 \|\vec{g}_1^i\|^2-|\langle\vec{h}_1^j,\vec{g}_1^i\rangle|^2) \notag\\
=&|I_{\rho}|({F_{j0}})|J_{\rho}|({E_{i0}})-\frac{1}{4}\left(\langle\langle\rho_1^{\widetilde{F}_{j0}}|,|\rho_1^{\widetilde{F}_{j0}}\rangle\rangle\langle\langle\rho_1^{\widetilde{E}_{i0}}|,|\rho_1^{\widetilde{E}_{i0}}\rangle\rangle-|\langle\langle\rho_1^{\widetilde{F}_{j0}}|,|\rho_1^{\widetilde{E}_{i0}}\rangle\rangle|^2\right)\notag\\
=&|I_{\rho}|({F_{j0}})|J_{\rho}|({E_{i0}})-\frac{1}{4}\langle 1|[\sqrt{\rho},F_{j0}]^{\dag}[\sqrt{\rho},F_{j0}]|1\rangle\langle1|\{\sqrt{\rho},E_{i0}\}^{\dag}\{\sqrt{\rho},E_{i0}\}|1\rangle \notag\\
+&\frac{1}{4}|\langle 1|[\sqrt{\rho},F_{j0}]^{\dag}\{\sqrt{\rho},E_{i0}\}|1\rangle|^2.
\end{align}
Thus we have $\widetilde{I}_{\rho}(\Psi)\widetilde{J}_{\rho}(\Phi)\geq \widetilde{I}_1$,
which is saturated if and only if $|\rho_1^{\widetilde{F}_{j0}}\rangle$ and $|\rho_1^{\widetilde{E}_{i0}}\rangle$ are linear dependent for arbitrary $i$ and $j$. Therefore, we have
$|U_{\rho}|(\Phi)|U_{\rho}|(\Psi)
=\sqrt{\widetilde{I}_{\rho}(\Phi)\widetilde{J}_{\rho}(\Psi)
\widetilde{I}_{\rho}(\Psi)\widetilde{J}_{\rho}(\Phi)}\\
\geq\sqrt{I_1\widetilde{I}_1}$. $\Box$

{\bf Theorem 4} For two arbitrary quantum channels with Kraus representations $\Phi(\rho)=\sum_{i}^NE_i\rho E_i^\dag$ and $\Psi(\rho)=\sum_{j}^NF_j\rho F_j^\dag$ on an $n$-dimensional Hilbert space $\mathcal{H}$, the following uncertainty relation holds,
\begin{align}\label{eq27}
|U_{\rho}|^2(\Psi)+|U_{\rho}|^2(\Phi)\geq&
\frac{1}{4}\sum_{ij}^N(|\langle[\sqrt{\rho},F_i],\{\sqrt{\rho},F_j\}\rangle|^2\\
&+\langle[\sqrt{\rho},E_i],[\sqrt{\rho},E_i]\rangle
(\langle\{\sqrt{\rho},E_j\},\{\sqrt{\rho},E_j\}\rangle\notag-4|\mathrm{Tr}(\rho E_j)|^2)).
\end{align}

{\it Proof} By using the Cauchy-Schwarz inequality, we have
\small{\begin{align*}
&|U_{\rho}|^2(\Psi)+|U_{\rho}|^2(\Phi)\notag\\
=&\widetilde{I}_{\rho}(\Psi)\widetilde{J}_{\rho}(\Psi)+\widetilde{I}_{\rho}(\Phi)\widetilde{J}_{\rho}(\Phi)\notag\\
=&\sum_{ij}^N\left[I_{\rho}(F_{i0})J_{\rho}(F_{j0})+I_{\rho}(E_{i0})J_{\rho}(E_{j0})\right]\notag\\
=&\frac{1}{4}\sum_{ij}^N[\langle[\sqrt{\rho},F_{i0}],[\sqrt{\rho},F_{i0}]\rangle\langle\{\sqrt{\rho},F_{j0}\},\{\sqrt{\rho},F_{j0}\}\rangle
+\langle[\sqrt{\rho},E_{i0}],[\sqrt{\rho},E_{i0}]\rangle\langle\{\sqrt{\rho},E_{j0}\},\{\sqrt{\rho},E_{j0}\}\rangle]\notag\\
\geq&\frac{1}{4}\sum_{ij}^N[|\langle[\sqrt{\rho},F_{i0}],\{\sqrt{\rho},F_{j0}\}\rangle|^2+\langle[\sqrt{\rho},E_{i0}],[\sqrt{\rho},E_{i0}]\rangle\langle\{\sqrt{\rho},E_{j0}\},\{\sqrt{\rho},E_{j0}\}\rangle].\notag\\
=&\frac{1}{4}\sum_{ij}^N[|\langle[\sqrt{\rho},F_i],\{\sqrt{\rho},F_j\}\rangle|^2+\langle[\sqrt{\rho},E_i],[\sqrt{\rho},E_i]\rangle(\langle\{\sqrt{\rho},E_j\},\{\sqrt{\rho},E_j\}\rangle-4|\mathrm{Tr}(\rho E_j)|^2)].
\end{align*}}
$\Box$

In the following examples, we compare our results with the existing ones. For
convenience, we denote by $LB$, $LB1$ and $LB2$ the right hand sides of (\ref{eq13}), (\ref{eq14}) and (\ref{eq27}), respectively.

{\bf Example 1} We consider the Werner state in the Hilbert space $C^4$,
$$
\rho_w=\left(\begin{array}{cccc}
         \frac{1}{3}\theta&0&0&0\\
         0&\frac{1}{6}(3-2\theta)&\frac{1}{6}(4\theta-3)&0\\
         0&\frac{1}{6}(4\theta-3)&\frac{1}{6}(3-2\theta)&0\\
         0&0&0&\frac{1}{3}\theta\\
         \end{array}
         \right),
$$
where $\theta\in[0,1]$. $\rho_w$ is separable when $\theta\in[0,\frac{1}{3}]$.

Let $\Phi$ and $\Psi$ be the quantum channels with the following Kraus operator $\{E_i\}$ and $\{F_j\}$, respectively,
\begin{equation*}
E_1=
\begin{pmatrix}
1& 0& 0& 0\\
0& \sqrt{1-p}& 0& 0\\
0& 0& 1& 0\\
0& 0& 0& \sqrt{1-p}
\end{pmatrix}, \quad
E_2=
\begin{pmatrix}
0& 0& 0& 0\\
0& \sqrt{p}& 0& 0\\
0& 0& 0& 0\\
0& 0& 0& \sqrt{p}
\end{pmatrix},
\end{equation*}
\begin{equation*}
F_1=
\begin{pmatrix}
\sqrt{1-q}& 0& 0& 0\\
0& 1& 0& 0\\
0& 0& \sqrt{1-q}& 0\\
0& 0& 0& 1
\end{pmatrix}, \quad
F_2=
\begin{pmatrix}
0& 0& 0& 0\\
\sqrt{q}& 0& 0& 0\\
0& 0& 0& 0\\
0& 0& \sqrt{q}& 0
\end{pmatrix},
\end{equation*}
where $0\leq p,q\leq1$. For $\theta=\frac{3}{4}$, the quantum state $\rho_{w}$ reduces to an incoherent state $\frac{1}{2}(|0\rangle\langle0|+|1\rangle\langle1|)\otimes\frac{1}{2}(|0\rangle\langle
0|+|1\rangle\langle1|)$, which is commutative with any operators. In this case, $\widetilde{I}_{\rho_w}(\Phi)=\sum_iI_{\rho_w}(E_{i0})=0$ and $\widetilde{I}_{\rho_w}(\Psi)=\sum_iI_{\rho_w}(F_{j0})=0$. Therefore, $|U_{\rho_w}|(\Phi)=|U_{\rho_w}|(\Psi)=0$.

For $\theta=1$, according to the Eqs. (\ref{eq23}), (\ref{eq25}) and the inequality (\ref{eq13}), we have
\begin{align*}
\sqrt{I_1\widetilde{I}_1}
=\frac{1}{72}\sqrt{(10\sqrt{1-p}+5p-10)(40(\sqrt{1-q}-1)+4q(1+4\sqrt{1-q})-3q^2)}
\end{align*}
and
\begin{align*}
LB=\frac{1}{4}\sum_{ij}|\mathrm{Tr}([F_j,{E_i}^{\dag}]\rho)|^2=0.
\end{align*}
According to the inequalities (\ref{eq14}) and (\ref{eq27}), we obtain
\begin{align*}
LB1=\frac{5}{72}(\sqrt{1-p}-1)^2(\sqrt{1-q}-1)^2
\end{align*}
and
\begin{align*}
LB2=\frac{5}{144}[(\sqrt{1-p}-1)^2+p]^2.
\end{align*}
Fig. 1 shows the relations between the lower bounds
$\sqrt{I_1\widetilde{I}_1}$ (the lower bound given in Theorem 3) and $LB$, and between $LB1$ and $LB2$, respectively.
\begin{figure}[ht]\centering
\subfigure[]
{\begin{minipage}[XuCong-Uncertainty-relation-7a]{0.49\linewidth}
\includegraphics[width=0.95\textwidth]{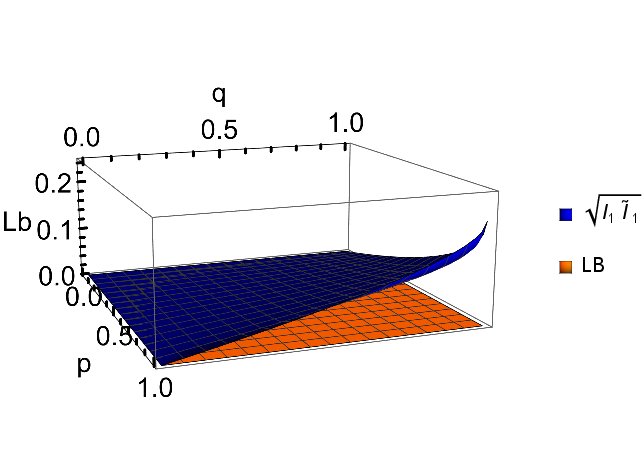}
\end{minipage}}
\subfigure[]
{\begin{minipage}[XuCong-Uncertainty-relation-7c]{0.47\linewidth}
\includegraphics[width=0.93\textwidth]{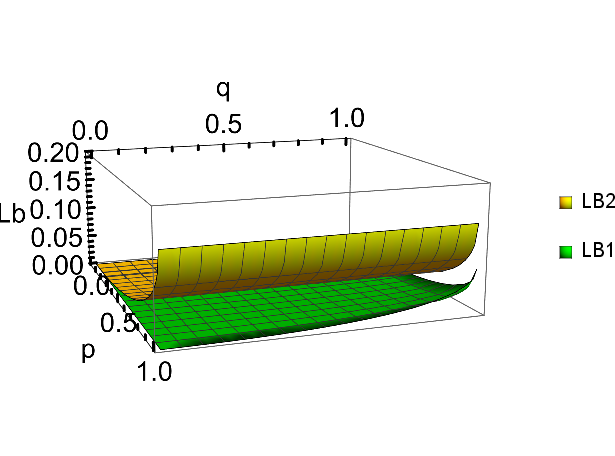}
\end{minipage}}
\caption{{ The blue, orange, green and yellow surfaces represent the lower bounds (Lb)
$\sqrt{I_1\widetilde{I}_1}$, $LB$, $LB1$ and $LB2$, respectively.  For the Werner state $\rho_w$ with $\theta=1$, (a) the comparison between $\sqrt{I_1\widetilde{I}_1}$ and $LB$; (b) the comparison between $LB1$ and $LB2$. \label{fig:Fig1}}}
\end{figure}

{\bf Example 2} Let us consider the following quantum state in the Hilbert space $C^4$,
$$
\rho_{\theta}=\left(\begin{array}{cccc}
         \frac{1}{4}&\frac{1}{4}(2\theta-1)&0&0\\
         \frac{1}{4}(2\theta-1)&\frac{1}{4}&0&0\\
         0&0&\frac{1}{4}&\frac{1}{4}(2\theta-1)\\
         0&0&\frac{1}{4}(2\theta-1)&\frac{1}{4}\\
         \end{array}
         \right),
$$
where $\theta\in[0,1]$. When $\theta=\frac{1}{2}$, similar to case of the Werner state $\rho_w$, for the quantum channels $\Phi$ and $\Psi$ in Example 1, we have $|U_{\rho_{\theta}}|(\Phi)=|U_{\rho_{\theta}}|(\Psi)=0$.

For $\theta=0$, according to the Eqs. (\ref{eq23}), (\ref{eq25}) and the inequality (\ref{eq13}), we have
\begin{align*}
\sqrt{I_1\widetilde{I}_1}
=\frac{1}{128}\sqrt{(2\sqrt{1-p}+p-2)(1800(\sqrt{1-q}-1)+60q(14+\sqrt{1-q})-q^2)}
\end{align*}
and
\begin{align*}
LB=\frac{1}{4}\sum_{ij}|\mathrm{Tr}([F_j,{E_i}^{\dag}]\rho)|^2=\frac{q}{8}(1-\sqrt{1-p}).
\end{align*}
According to the inequalities (\ref{eq14}) and (\ref{eq27}), we obtain
\begin{align*}
LB1=\frac{1}{8}(1-\sqrt{1-p})(1-\sqrt{1-q}+\sqrt{pq})^2
\end{align*}
and
\begin{align*}
LB2=\frac{1}{16}[(1-\sqrt{1-p})^4+p^2+2|p(p-2+2\sqrt{1-p})+q(q-2+2\sqrt{1-q})|].
\end{align*}
\begin{figure}[ht]\centering
\subfigure[]
{\begin{minipage}[XuCong-Uncertainty-relation-7b]{0.49\linewidth}
\includegraphics[width=0.95\textwidth]{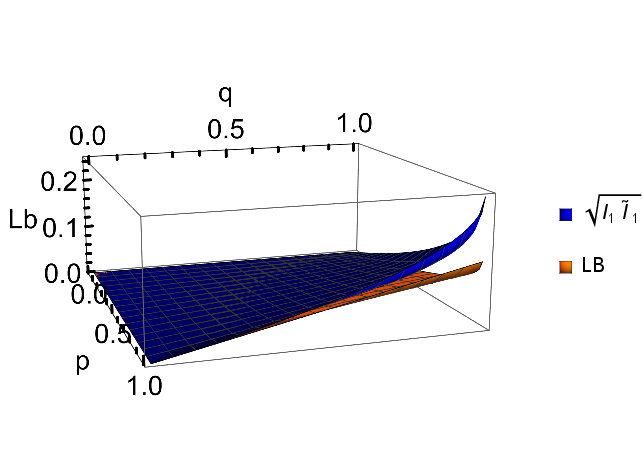}
\end{minipage}}
\subfigure[]
{\begin{minipage}[XuCong-Uncertainty-relation-7d]{0.47\linewidth}
\includegraphics[width=0.93\textwidth]{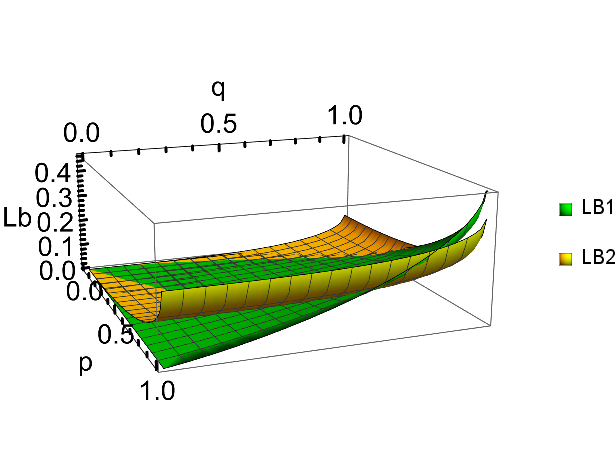}
\end{minipage}}
\caption{{ The blue, orange, green and yellow surfaces represent the lower bounds (Lb)
$\sqrt{I_1\widetilde{I}_1}$, $LB$, $LB1$ and $LB2$, respectively. For quantum  state $\rho_\theta$, $\theta=0$, (a) the comparison of the lower bound between $\sqrt{I_1\widetilde{I}_1}$ and $LB$, (b) the comparison of the lower bound between $LB1$ and $LB2$. \label{fig:Fig2}}}
\end{figure}

From Fig. $1$ $(a)$ and Fig. $2$ $(a)$, we see that the blue surface covers fully the orange one for arbitrary $0\leq p,q\leq 1$. This means that the lower bound of our Theorem 3 is tighter than $LB$. From Fig. $1$ $(b)$, we find that the yellow surface covers the green one. However, from Fig. $2$ $(b)$ it is obvious that the yellow surface covers the green one only for some cases. This implies that the lower bound of our Theorem 4 is tighter than $LB1$ for some classes of states.

From Fig. $1$ $(a)$ and Fig. $2$ $(a)$, we see that all the lower bounds are equal to zero when $p=0$ or $q=0$. In this case, the quantum channels $\Phi$ or $\Psi$ reduces to the identity channel. From Fig. $1$ $(b)$, we can obtain a similar conclusion for $p=0$.

%

\noindent {\bf 4. Conclusions}

We have explored the product form uncertainty relations of quantum channels in terms of the symmetrized $\rho$-absolute variance, which generalizes the results in Ref. \cite{DD1} from operators to quantum channels. By using Cauchy-Schwarz inequality, we have also presented the product and summation form uncertainty relations of quantum channels in terms of the quantity $|U_{\rho}|(\Phi)$. By explicit examples, we have shown that our lower bounds are tighter than the existing ones. Our results may highlight further researches on uncertainty relations of quantum channels characterized in other ways.
\vskip0.1in

\noindent

\subsubsection*{Acknowledgements}
This work was supported by National Natural Science Foundation of
China (Grant Nos. 12161056, 12075159, 12171044); the
specific research fund of the Innovation Platform for Academicians of Hainan Province under Grant No. YSPTZX202215; and Changsha University of Science
and Technology (Grant No. 097000303923).

\subsubsection*{Conflict of interest}
\small {The authors declare that they have no conflict of interest.}



\end{document}